\documentclass[10pt,conference]{IEEEtran}
\IEEEoverridecommandlockouts
\nocite{IEEEexample:BSTcontrol}
\usepackage{cite}
\usepackage{amsmath,amssymb,amsfonts}
\usepackage{textcomp}
\usepackage{xcolor}
\usepackage{geometry}
\usepackage[utf8]{inputenc} %
\usepackage[T1]{fontenc}    %
\usepackage{pifont}
\usepackage[normalem]{ulem}
\usepackage[colorlinks,
    linkcolor=red,citecolor=citecolor,urlcolor=blue]{hyperref}       %

\usepackage{graphicx}
\usepackage{xcolor}
\usepackage{todonotes}
\usepackage[table,xcdraw]{xcolor}
\usepackage{amssymb}
\usepackage{threeparttable}
\usepackage{multirow}
\usepackage{amsmath}
\usepackage{bm}
\usepackage{float}
\usepackage{amsthm}
\usepackage{soul}
\usepackage[noend]{algpseudocode}
\usepackage{enumitem} %
\usepackage[subrefformat=parens,labelformat=parens]{subfig}

\usepackage{xspace}

\usepackage{multirow}

\definecolor{citecolor}{RGB}{34,139,34}
\definecolor{mydarkblue}{rgb}{0,0.08,1}
\definecolor{mydarkgreen}{rgb}{0.02,0.6,0.02}
\definecolor{mydarkred}{rgb}{0.8,0.02,0.02}
\definecolor{mydarkorange}{rgb}{0.40,0.2,0.02}
\definecolor{mypurple}{RGB}{111,0,255}
\definecolor{myred}{rgb}{1.0,0.0,0.0}
\definecolor{mygold}{rgb}{0.75,0.6,0.12}
\definecolor{myblue}{rgb}{0,0.2,0.8}
\definecolor{mydarkgray}{rgb}{0.,0.2,0.2}

\definecolor{lightred}{RGB}{255,235,235}
\definecolor{lightgreen}{RGB}{235,255,235}
\definecolor{lightblue}{RGB}{235,235,255}
\definecolor{lightcyan}{RGB}{235,255,255}
\definecolor{lightmagenta}{RGB}{255,235,255}
\definecolor{lightyellow}{RGB}{255,255,235}

\definecolor{qxkcolor}{RGB}{215,235,255}
\definecolor{softmaxcolor}{RGB}{230,235,255}
\definecolor{probxvcolor}{RGB}{255,255,235}

\definecolor{topkcolor}{RGB}{255,235,235}
\definecolor{zecolor}{RGB}{255,255,235}
\definecolor{dynacolor}{RGB}{235,255,255}

\definecolor{reviewcolor}{RGB}{0,0,200}

\theoremstyle{plain}

\theoremstyle{definition}

\algdef{SE}[DOWHILE]{Do}{doWhile}{\algorithmicdo}[1]{\algorithmicwhile\ #1}%

\newcommand{\squishlist}{
 \begin{list}{$\bullet$}
  { \setlength{\itemsep}{0pt}
     \setlength{\parsep}{3pt}
     \setlength{\topsep}{3pt}
     \setlength{\partopsep}{0pt}
     \setlength{\leftmargin}{1.5em}
     \setlength{\labelwidth}{1em}
     \setlength{\labelsep}{0.5em} } }
     
\newcommand{\squishend}{
  \end{list}  }

\def\BibTeX{{\rm B\kern-.05em{\sc i\kern-.025em b}\kern-.08em
    T\kern-.1667em\lower.7ex\hbox{E}\kern-.125emX}}

\begin{document}

\title{Toward Lifelong-Sustainable Electronic-Photonic AI Systems via Extreme Efficiency, Reconfigurability, and Robustness
\vspace{-10pt}
}

\author{
Ziang Yin,
Hongjian Zhou, 
Chetan Choppali Sudarshan,
Vidya Chhabria,
Jiaqi Gu$^{\dagger}$\\
Arizona State University \\ {\small $\dagger$jiaqigu@asu.edu}
\vspace{-10pt}
}

\maketitle
\begin{abstract}

The relentless growth of large-scale artificial intelligence (AI) has created unprecedented demand for computational power, straining the energy, bandwidth, and scaling limits of conventional electronic platforms. 
Electronic-photonic integrated circuits (EPICs) have emerged as a compelling platform for next-generation AI systems, offering inherent advantages in ultra-high bandwidth, low latency, and energy efficiency for computing and interconnection. 
Beyond performance, EPICs also hold unique promises for sustainability. 
Fabricated in relaxed process nodes with fewer metal layers and lower defect densities, photonic devices naturally reduce embodied carbon footprint (CFP) compared to advanced digital electronic integrated circuits, while delivering orders-of-magnitude higher computing performance and interconnect bandwidth.
To further advance the sustainability of photonic AI systems, we explore how electronic–photonic design automation (EPDA) and cross-layer co-design methodologies can amplify these inherent benefits. 
We present how advanced EPDA tools enable more compact layout generation, reducing both chip area and metal layer usage. 
We will also demonstrate how cross-layer device-circuit-architecture co-design unlocks new sustainability gains for photonic hardware: ultra-compact photonic circuit designs that minimize chip area cost, reconfigurable hardware topology that adapts to evolving AI workloads, and intelligent resilience mechanisms that prolong lifetime by tolerating variations and faults.
By uniting intrinsic photonic efficiency with EPDA- and co-design-driven gains in area efficiency, reconfigurability, and robustness, we outline a vision for lifelong-sustainable electronic–photonic AI systems. 
This perspective highlights how EPIC AI systems can simultaneously meet the performance demands of modern AI and the urgent imperative for sustainable computing.

\end{abstract}

\section{Introduction}
The rapid growth of modern computing, spanning mobile devices to massive AI data centers, has created an unprecedented demand for computational power. 
Training trillion-parameter AI models and operating large-scale data centers require massive energy and hardware resources, raising urgent concerns about sustainability.
The information and communication technology (ICT) sector already contributes over 2–3$\%$ of global carbon emissions~\cite{ai_sus_prob,ict-carbon-report}, a share comparable to the aviation industry, and this contribution is projected to increase if current trends continue.

Historically, the semiconductor industry has responded to rising computational demand through aggressive technology scaling and efficiency improvements to reduce operational power consumption.
However, this progress has introduced a critical imbalance: while operational energy per operation has declined, the \emph{embodied carbon cost of design, fabrication, and packaging has grown to dominate lifecycle carbon emissions}~\cite{ai_sus_prob}.
The impact of this imbalance is evident in advanced hardware manufacturing~\cite{sudarshan2023ecochip}. 
Fabricating a single state-of-the-art (SoTA) digital processor can release hundreds of kilograms of $CO_2$, with extreme ultraviolet (EUV) lithography consuming nearly an order of magnitude more energy than older deep ultraviolet (DUV) processes~\cite{imec-dtco}. 
As a result, even though each chip generation is more energy-efficient in use, its sustainability often worsens. This paradox has elevated sustainability from a secondary concern to a primary design constraint.

Addressing this challenge requires a clear definition of what determines a chip’s total carbon footprint (CFP).
Recent efforts such as ACT~\cite{act} and ECO-CHIP~\cite{sudarshan2023ecochip} show that sustainability cannot be characterized by power efficiency alone. 
Instead, CFP comprises two dominant components: \textbf{(1) embodied carbon}, arising from design, fabrication, packaging, and material waste, and \textbf{(2) operational carbon}, determined by energy consumption over the chip’s lifetime. 
For many modern devices, embodied costs already outweigh operational costs, revealing that optimizing for energy efficiency alone is insufficient. 
To address this, researchers have proposed new carbon-aware metrics, such as the carbon-delay product (CDP) and the carbon footprint to throughput ratio (CFP/FPS), which extend the traditional performance-power–area (PPA) paradigm to explicitly incorporate carbon considerations.

These efforts expose the limits of electronic integrated circuits (EICs). 
As Moore’s law slows, the \emph{embodied footprint of advanced electronic integrated circuits (EICs) has ballooned}: EUV photolithography, dozens of metal layers, and ever-denser transistors drive up CFP, even as device efficiency improves. 

Photonic integrated circuits (PICs) have emerged as a compelling technology for AI acceleration.
By harnessing the unique properties of light for computing and communication, photonic AI hardware (optical neural networks, ONNs) delivers ultra-high bandwidth, low latency, and exceptional energy efficiency per operation~\cite{NP_NATURE2017_Shen, NP_Nature2020_Wetzstein, NP_NaturePhotonics2021_Shastri, NP_Nature2021_Xu, NP_Nature2021_Feldmann, NP_Optica2025_Ning, NP_JLT20254_Ning,NP_Nature2025_Kalinin, NP_Nature2025_Hua, NP_Nature2025_Ahmed}.
Photonic interconnects and switches further enable low-latency, high-bandwidth communication between chips and racks, outperforming electrical networks and already proving indispensable in large-scale AI clusters such as Google’s TPUv4~\cite{TPUV4, NP_OFT2020_Cheng}. 

\begin{figure}
    \centering
    \includegraphics[width=1\columnwidth]{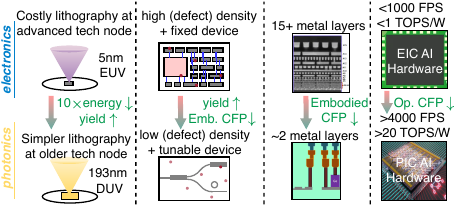}
    \vspace{-10pt}
    \caption{Sustainability advantages of photonic technology.}
    \label{fig:Illustration_Photonics}
    \vspace{-13pt}
\end{figure}

Photonics also carries intrinsic advantages in environmental sustainability~\cite{fayza2024photonicssustainablecomputing}, as shown in Fig.~\ref{fig:Illustration_Photonics}.

\ding{202}~PIC fabrication leverages relaxed technology nodes and DUV, while requiring far fewer metal layers and exhibiting lower defect densities due to the sparse layout. 
Larger device geometries translate into higher yield, reduced material waste, and significantly lower CFP per chip. 
Moreover, active PICs typically require only two metal layers for routing, far fewer than the 15+ layers in modern digital EICs, significantly reducing fabrication steps, mask count, and associated emissions. 

\ding{203}~The sustainability benefits extend further when considering system-level amortization of embodied CFP.
Thanks to their high runtime reconfigurability, PIC-based systems can adapt to diverse and evolving AI workloads, preventing premature obsolescence and extending system lifetimes.
Robust and tunable photonic devices mitigate fabrication variations and thermal drift, while reconfigurable interconnects can bypass defective or noisy components at runtime.

\ding{204}~Finally, the operational carbon footprint of photonic AI systems is drastically lower than their electronic counterparts, owing to unmatched energy efficiency per operation and speed-of-light interconnects.

In this work, we explore how these \textbf{intrinsic strengths of photonics can be amplified through electronic–photonic design automation (EPDA) and cross-layer co-design methodologies}. 
We argue that the combined pillars of extreme efficiency, reconfigurability, and robustness pave the way toward lifelong-sustainable electronic–photonic AI systems that simultaneously deliver breakthrough AI performance and address the urgent imperative of sustainable computing.

\vspace{-5pt}
\section{Toward Sustainable EPIC via Extreme Efficiency, Reconfigurability, and Robustness Cross-Layer Co-Design}
\vspace{-5pt}
Electronic-photonic integrated circuits (EPICs) combine the density and maturity of electronics with the unique physical properties of photonics, enabling unprecedented levels of efficiency, adaptability, and reliability. We highlight three cross-layer co-design pillars, \ding{202}~\textbf{extreme area and power efficiency}, \ding{203}~\textbf{reconfigurability}, and \ding{204} \textbf{robustness}, that are essential for achieving both breakthrough performance and long-term sustainability.
In this section, we illustrate these pillars through \emph{case studies} from recent photonic accelerator designs, drawing broader insights into how \emph{cross-layer co-design methodology} can amplify the intrinsic advantages of photonics.

\subsection{Toward Sustainable EPIC via Area and Power Optimization}
\noindent\underline{\textbf{Chip Area Compression}}.~
Chip area directly impacts both embodied carbon footprint (CFP) and system cost. 
Photonic AI chip area optimization must balance compactness with robustness against crosstalk and thermal variation, making area optimization a cross-layer challenge.

We discuss a prior photonic AI accelerator \texttt{SCATTER}~\cite{NP_ICCAD2024_Yin} as a case study to demonstrate how progressive co-design across devices, circuits, and algorithms can deliver orders-of-magnitude improvements in power-area efficiency.
\begin{figure}
    \centering
    \includegraphics[width=\columnwidth]{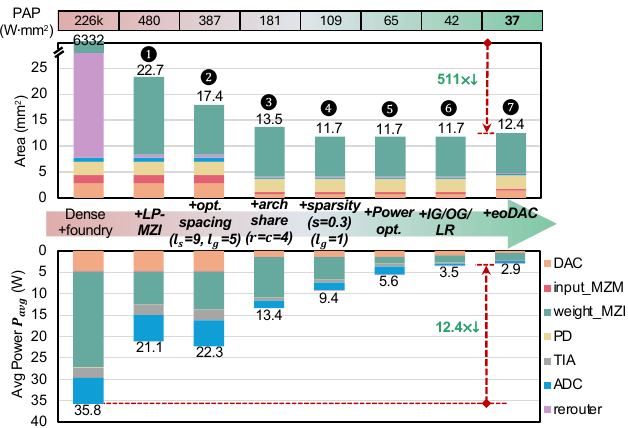}
    \vspace{-15pt}
    \caption{Significant power-area-product reduction can be achieved by progressively adding our proposed cross-layer optimization and algorithmic-circuit co-sparsity techniques~\cite{NP_ICCAD2024_Yin}.
    }
    \label{fig:ProgressPowerAreaOpt}
    \vspace{-15pt}
\end{figure}

The progressive area optimization that leads to 511$\times$ area reduction is demonstrated in Fig.~\ref{fig:ProgressPowerAreaOpt}.

\ding{202}~\textbf{Device-Level Optimization} --
Replacing foundry Mach-Zehnder Interferometer (MZI) switches with customized compact low-power MZIs (LP-MZIs) reduced chip area by nearly 279$\times$ while lowering power by 41\%.

\ding{203}~\textbf{Layout-Level Densification} -- 
Device spacing is optimized to balance area, power consumption, and crosstalk area, which enables another 23.3\% area reduction.

\ding{204}~\textbf{Architecture-Level Hardware Sharing} --
Architectural sharing of input modulators and readout circuitry amortized peripheral DAC/ADC overheads, yielding an additional 40\% area reduction and 22\% power savings.

\ding{205}~\textbf{Algorithm-Circuit Co-Sparsity} -- 
Algorithm-circuit co-sparsity mostly eliminates thermal crosstalk effects across adjacent thermo-optic MZI weighting units, allowing extremely tight device packing while preserving accuracy.

\noindent\textit{\emph{Insight}}.~
Cross-layer co-optimization can overcome fundamental photonic scaling barriers. 
With co-optimized chip area, thermal robustness, and algorithm, denser layouts become feasible.

\noindent\underline{\textbf{Operational Energy Efficiency Optimization}}.~
While embodied carbon dominates the sustainability profile of advanced hardware, operational energy remains a non-negligible contributor, particularly in large-scale AI clusters.
Optimizing energy efficiency is therefore critical to reduce lifetime carbon footprint while simultaneously enabling higher throughput per watt. 
In this section, we highlight three representative case studies that demonstrate complementary strategies for energy optimization at different design levels.

\ding{202}~\textbf{Device-Circuit-Algorithm Co-Design \texttt{SCATTER}} --
In Fig.~\ref{fig:ProgressPowerAreaOpt}, \texttt{SCATTER} integrates workload-aware structured sparsity with topology-reconfigurable photonic circuit design to aggressively gate unused devices and redistribute optical power. 
This not only reduces static and dynamic power consumption but also improves signal-to-noise ratio, enabling stable inference at significantly lower energy budgets. 
It is also equipped with novel electronic-optic DACs that realize high programming precision with low power consumption.
The result is a 12.4$\times$ reduction of power consumption, illustrating how cross-layer co-design can directly lower energy per operation.

\ding{203}~\textbf{Electronic-Photonic Architecture Optimization} --
Besides the active device power and laser power, another key barrier to photonic AI accelerators is the high overhead of electrical–optical (E–O) and optical–electrical (O–E) conversions.
LighteningTransformer~\cite{NP_HPCA2024_Zhu} and TeMPO~\cite{NP_JAP2024_Zhang} both combine wavelength-division or time-division multiplexing (WDM/TDM) for higher parallelism with a crossbar topology for operand sharing.
To further cut conversion overheads, the architecture introduces two system-level optimizations: 
(1) \emph{optical interconnect broadcast}, which shares common operands across cores by leveraging the natural broadcast capability of optics; and 
(2) \emph{analog-domain temporal accumulation}, which integrates partial sums directly in the optical domain, reducing ADC frequency and power while maintaining digital-level accuracy. 
Across Transformer benchmarks across vision and language tasks, LightningTransformer achieves over \textbf{2.6$\times$ energy savings} and 12$\times$ latency reduction compared to prior photonic designs, and delivers 2–3 orders of magnitude lower energy–delay product than SoTA electronic accelerators.

\ding{204}~\textbf{Hybrid Optical/Photonic AI Hardware: \texttt{CHORD}} --
Integrated photonic accelerators provide reconfigurability and precision, while free-space optical diffractive systems excel at ultra-fast, passive spatial processing.
\texttt{CHORD}~\cite{NP_DAC2025_Yin_CHORD} unites these strengths into a hybrid architecture: diffractive optics serve as a high-throughput front end for large-scale feature sensing, transformation, and compression~\cite{NN_PhysRevLett19, NP_SciAdv24_Wei, NP_OPTICA24_Peng, NN_CHEN2021, NP_Science18_Lin,NN_Optical23_Anderson}, while a lightweight photonic tensor core backend handles the remaining programmable tasks. 
This division of labor offloads most of the computation onto passive optical layers, dramatically relieving the burden on the photonic backend.
Architectural simulation shows that CHORD achieves 74$\times$ faster inference and 194$\times$ lower energy than a digital CNN baseline on an NVIDIA A6000 GPU. 
The diffractive layers introduce negligible latency (<1 ns equivalent), allowing the photonic backend to operate with reduced scale and power demands. 
In aggregate, \texttt{CHORD} achieves unprecedented performance density, reaching 4327 TOPS/W and 1032 TOPS/mm².

\begin{figure}
    \centering
    \includegraphics[width=0.9\columnwidth]{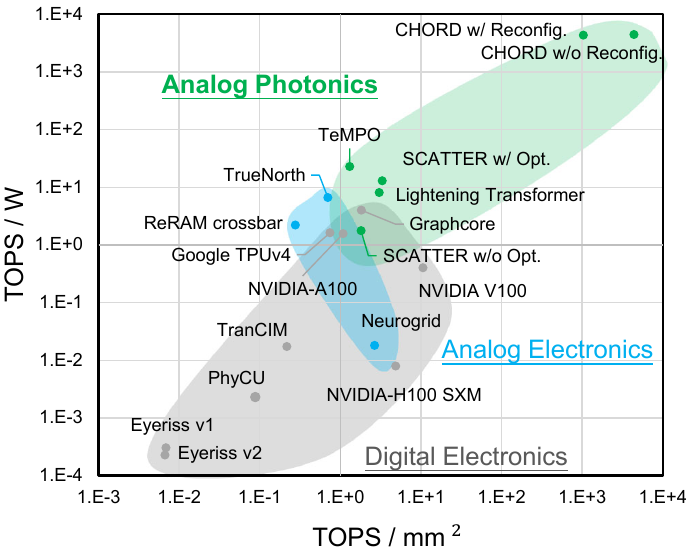}
    \caption{Efficiency and density Pareto front comparison of different AI accelerators, including electronic ASICs, GPUs/TPUs, and SoTA EPIC AI hardware~\cite{Chen2017EyerissJSSC, Chen2019EyerissV2, Ju2024PhyCU, NP_JAP2024_Zhang, NP_ICCAD2024_Yin, NP_HPCA2024_Zhu, TPUV4, Tu2023TranCIM, NVIDIA2022H100Whitepaper, NVIDIA2022H100PCIeBrief, NP_DAC25_Yin}}
    \label{fig:ParetoFrontier}
    \vspace{-10pt}
\end{figure}

\noindent\textit{\underline{Insight}}.~
Together, these case studies demonstrate that operational energy efficiency emerges from coordinated innovations across the design stack:
(1) Device-circuit co-design that balances robustness, area, and power. 
(2) Architectural optimization that leverages the photonic-unique dataflow can significantly reduce the dominant E-O/O-E conversion cost via optical broadcast and analog-domain accumulation.
(3) Hybrid optic–photonic composition exploits passive optics for ultra-parallel spatial transforms, relieving backend pressure and achieving record throughput-per-watt.
By aligning optimization at every layer, EPIC-based AI systems can deliver orders-of-magnitude improvements in energy–delay product, directly lowering operational carbon footprint while sustaining accuracy and scalability.
Figure~\ref{fig:ParetoFrontier} shows that co-designed photonic AI hardware achieves a better Pareto frontier in energy efficiency and compute density compared to SoTA electronic accelerators.

\vspace{-5pt}
\subsection{Toward Sustainable EPIC via Extreme Versatility and Reconfigurability}
\label{subsec:tsevvar}
Beyond area/energy efficiency, \textit{long-term sustainability also depends on reconfigurability and versatility}.
AI workloads evolve rapidly, from convolutional networks, to Transformers, to emerging multi-modal architectures, and fixed-function accelerators risk premature obsolescence when hardware cannot adapt. 
In this section, we present reconfigurable and multifunctional EPIC AI accelerator designs that can amortize their embodied carbon footprint across diverse tasks, extending their practical lifecycle and improving hardware utilization.

\begin{figure}
    \centering
    \includegraphics[width=0.9\columnwidth]{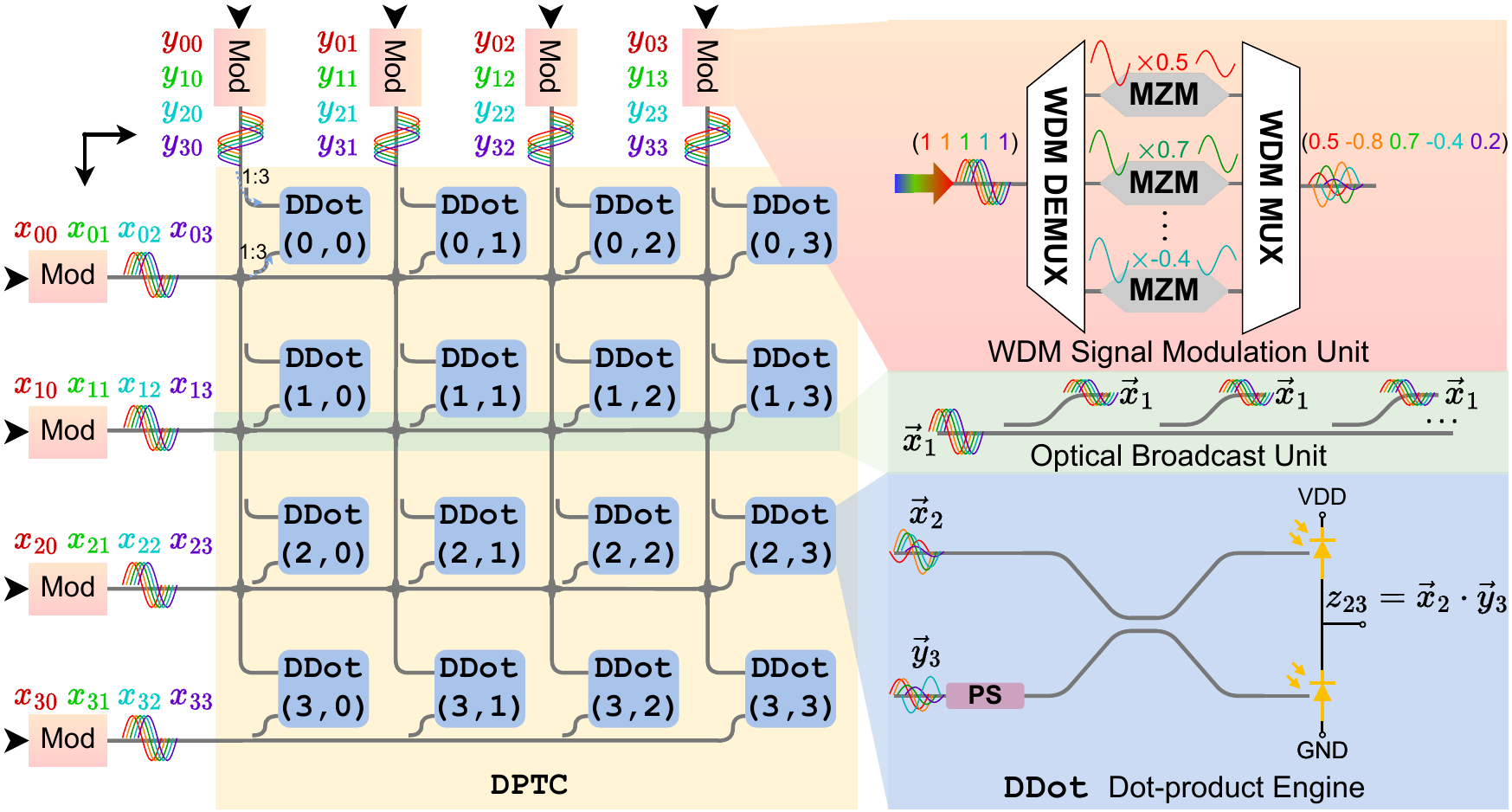}
    \vspace{-8pt}
    \caption{LighteningTransformer~\cite{NP_HPCA2024_Zhu} with dynamic tensor cores for versatile AI inference acceleration.}
    \label{fig:lighteningTransformer}
    \vspace{-10pt}
\end{figure}

\textbf{Versatile Dynamic Photonic Tensor Accelerator} --
LighteningTransformer~\cite{NP_HPCA2024_Zhu} introduces a dynamic tensor core (DPTC), as shown in Fig.~\ref{fig:lighteningTransformer}. 
\textit{Both operands of the DPTC are optically encoded at high speed}, allowing weights and activations to be flexibly updated at runtime. 
This enables the DPTC to efficiently execute not only static operations such as convolutions and linear layers, but also highly dynamic ones such as self-attention in large language models.
This versatility is critical for modern AI workloads.
The key architectural insight is that versatility (supporting multiple AI primitives) and dynamic reconfigurability can be unified within a single photonic tensor core design, turning what was once a fixed-function photonic block into a \emph{multifunctional, sustainable accelerator}.

\begin{figure}
    \centering
    \includegraphics[width=0.95\columnwidth]{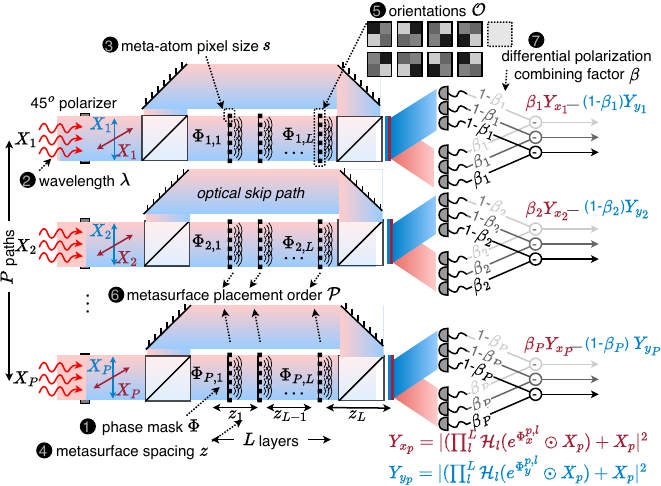}
    \caption{Composable Hybrid Optical Reconfigurable Diffractive Framework (CHORD)~\cite{NP_DAC2025_Yin_CHORD}. Parameters \ding{202}–\ding{208} represent system-level reconfigurable variables that expand expressivity exponentially beyond traditional phase-mask-only systems.}
    \label{fig:chord}
    \vspace{-5pt}
\end{figure}

\textbf{System-Level Multi-Dimensional Reconfigurability} --
\texttt{CHORD} exemplifies the principle of sustainability through reconfigurability by introducing a physically composable diffractive framework integrated with programmable photonic cores~\cite{NP_DAC2025_Yin_CHORD} as shown in Fig.~\ref{fig:chord}.
Unlike static diffractive optical neural networks that are constrained to the functions encoded at fabrication, \texttt{CHORD} unlocks \emph{exponentially higher expressivity} through multi-\textit{dimensional system-level tunability}, including wavelength, polarization, metasurface spacing, orientation, and placement order, as well as \textit{lightweight photonic tensor core weights}.
This system-level reconfigurability vastly expands the functional space of a single hardware instance, supporting rapid adaptation to diverse applications without the need for re-manufacturing. 
Table~\ref{tab:chord_result} shows that even when the optical diffractive hardware is fabricated and fixed, by adapting end-to-end system parameter training, \texttt{CHORD} can adapt to various new tasks, even for partial differential equation solving.

\begin{table}[t]
\centering
\caption{Evaluate \texttt{CHORD}~\cite{NP_DAC2025_Yin_CHORD} on pre-fab initial tasks and post-fab adapted tasks.
A 2-layer digital CNN (C32K3)$_{\times 2}$ is shown as a reference.
For classification, test Acc. is shown.
For Darcy/NavierStokes, MSE is shown.}
\label{tab:chord_result}
\vspace{-5pt}
\resizebox{1\columnwidth}{!}{
\begin{tabular}{l|c|c|c}
\hline
\multicolumn{1}{c|}{Tasks} & \begin{tabular}[c]{@{}c@{}}Digital 2-Layer CNN \\ C32K3-C32K3\end{tabular} & \begin{tabular}[c]{@{}c@{}}Pre-fab\\ Initial Training\\ $P$=4, $L$=4, $\alpha$, $\Phi$, \\ $\beta$, $Skip$, $s$, $\lambda$\end{tabular} & \begin{tabular}[c]{@{}c@{}}Post-fab\\ Task Adaptation\\ $P$=4, $L$=4, $\alpha$, $\beta$, $Skip$, \\$\lambda$, $\mathcal{O}$, $\mathcal{P}$\end{tabular} \\ \hline
FMNIST           & 0.9122                                                                       & \cellcolor[HTML]{E6E6E6}0.9014                                                                                        & \cellcolor[HTML]{E6E6E6}-                                                                                                               \\
Quickdraw        & 0.8978                                                                       & \cellcolor[HTML]{E6E6E6}0.8646                                                                                        & \cellcolor[HTML]{E6E6E6}0.7992 (adapt from FMNIST)                                                                                                         \\ \hline
CIFAR10          & 0.7633                                                                       & \cellcolor[HTML]{E6E6E6}0.6182                                                                                        & \cellcolor[HTML]{E6E6E6}-                                                                                                               \\
CIFAR100         & 0.4347                                                                       & \cellcolor[HTML]{E6E6E6}0.3815                                                                                        & \cellcolor[HTML]{E6E6E6}0.2513 (adapt from CIFAR10)                                                                                                         \\ \hline
Darcy           & 0.0466                                                                       & \cellcolor[HTML]{E6E6E6}0.055                                                                                         & \cellcolor[HTML]{E6E6E6}-                                                                                                               \\
NavierStokes    & 0.1085                                                                       & \cellcolor[HTML]{E6E6E6}0.1007                                                                                        & \cellcolor[HTML]{E6E6E6}0.1753 (adapt from Darcy)                                                                                                        \\ \hline
\end{tabular}
}
\vspace{-8pt}
\end{table}

\noindent\textit{\underline{Insight}}:~
Lightning-Transformer and CHORD highlight complementary paths toward versatile and sustainable EPICs.
At the operand level, we need to ensure that the hardware can adapt as dominant AI models evolve, which prevents obsolescence at the application level.
At the system level, we can explore the vast new degrees of reconfigurability beyond the optical hardware substrate itself, enabling fabricated substrates to be repurposed across entirely different tasks, amortizing embodied carbon footprint through prolonged reuse and adaptability.

\subsection{Toward Sustainable EPIC via Robustness Optimization}
\label{subsec:tsevro}
While efficiency and reconfigurability highlight the advantages of EPICs, their long-term endurance ultimately depends on robustness to device non-idealities and aging effects. 
Thermal drift, fabrication variation, crosstalk, and device wear-out can accumulate over time, degrading accuracy and shortening lifetime.
Here, we present robustness-driven optimization techniques to reinforce the lifetime of photonic AI hardware.

\begin{figure}
    \centering
    \includegraphics[width=0.95\columnwidth]{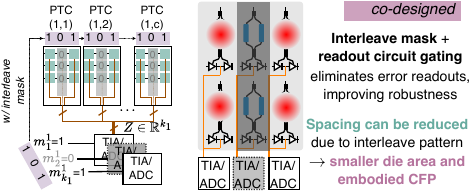}
    \caption{
    Customized PTC design \texttt{SCATTER}~\cite{NP_ICCAD2024_Yin} with thermal-aware, area-efficient circuit design and dynamic signal redistribution/gating enables compact layout without crosstalk issue.}
    \label{fig:scatterps}
    \vspace{-10pt}
\end{figure}

\ding{202}~\textbf{Built-in Thermal Robustness in Circuit Design} --
\texttt{SCATTER}~\cite{NP_ICCAD2024_Yin} integrates robustness directly into its architecture by co-optimizing workload sparsity and circuit sparsity. 
As shown in Fig.~\ref{fig:scatterps}, an interleaved sparsity pattern allows denser device packing while \emph{alternating active and inactive channels suppresses thermal crosstalk}. 
In-situ light redistribution and output gating further \emph{reallocate power} to active channels, simultaneously improving signal-to-noise ratio and reducing power. Together, these mechanisms enable compact layouts that are both robust to variations and more sustainable through lower embodied and operational costs.

\ding{203}~\textbf{Dynamic On-Chip Remediation Against Thermal Drift} --
Beyond robust hardware design, on-chip dynamic calibration and in-situ training are also effective approaches to resolve the robustness issue of photonic accelerators~\cite{NP_NeurIPS2021_Gu}.
To address temporally drifting noise and thermal variation, the DOCTOR~\cite{NP_JLT2024_Lu} framework is a representative SoTA method that introduces lightweight, in-situ calibration to resume photonic AI hardware accuracy under hardware nonideality.
By probing chip states and applying salience-aware sparse weight calibration, DOCTOR selectively corrects critical weights with negligible runtime overhead. 
In parallel, variation-aware tile remapping reassigns sensitive workloads to more stable cores, leveraging the nonuniform noise distribution across devices. 
With DOCTOR-assisted calibration, photonic AI accelerators, even with thermal-sensitive microring resonators (MRRs), show sustained accuracy recovery with only 0.1–5\% cycle overhead and 34\% higher accuracy than prior on-chip training, ensuring that photonic accelerators can \emph{remain reliable with extended lifetime under real-world nonideal, drifting environments}.

\ding{204}~\textbf{Endurance-Enhanced Photonic In-Memory Computing} --
For photonic in-memory tensor cores based on non-volatile phase-change materials (PCMs), limited write endurance poses a major lifetime bottleneck.
ELight~\cite{NP_TCAD2022_Gu_ELight} introduces a framework that introduces Write-aware training, boosting weight matrix cross-block similarity to minimize redundant PCM cell reprogramming.
It also employs fine-grained post-training weight reordering to further reduce PCM write operations.
A group-wise remapping technique is proposed to tolerate aged PCM cells by reassigning weight rows adaptively.
Together, these techniques cut write operations and dynamic energy by over 20$\times$ while maintaining accuracy, effectively prolonging device lifetime and mitigating post-aging degradation.

\noindent\textit{\underline{Insight}}.~
These case studies demonstrate that robustness in EPICs is not an afterthought, but a defining prerequisite for sustainability. 
By combining robust circuit designs (e.g., sparsity- and crosstalk-aware designs), with dynamic in-situ calibration against temporal drift and aging-aware weight programming and remapping, EPIC AI systems can maintain accuracy under variability, adapt to changing conditions, and endure significantly longer lifetimes to amortize the embodied CFP.

\newcommand{\nameplacer}{\texttt{Apollo}\xspace}
\newcommand{\namerouter}{\texttt{LiDAR}\xspace}
\section{Toward Sustainable EPIC via Automated Electronic-Photonic Design Automation (EPDA)}
Designing efficient EPIC AI systems requires more than device and architecture innovation, it also depends on \textbf{design automation}.
Today, the physical design of photonic integrated circuits (PICs) remains largely manual, requiring weeks of expert effort to \textit{satisfy design rules while balancing die area, routing layer usage, and circuit robustness}.
This slow, heuristic-driven process makes it infeasible to explore large design spaces or optimize layouts for sustainability. 
More importantly, manual layouts tend to be \emph{conservative, inflating chip area, increasing the number of routing layers, and struggling to balance CFP and other key specifications}, ultimately raising the embodied carbon footprint (CFP) of manufacturing.

Automated EPDA tools can fundamentally change this trajectory.
By systematically optimizing placement and routing solutions, automated engines can generate \textbf{more compact layouts for large-scale EPICs and consume fewer routing layers while still honoring design rules, preserving robustness, and satisfying various design constraints}. 
This directly translates into lower embodied carbon cost by reducing die size, lithography masks, and metal usage. 
At the same time, automation drastically reduces design time, allowing rapid evaluation of CFP vs. chip performance trade-offs early in the \emph{CFP-aware EPIC physical design cycle}.

\begin{figure}
    \centering
    \includegraphics[width=0.9\columnwidth]{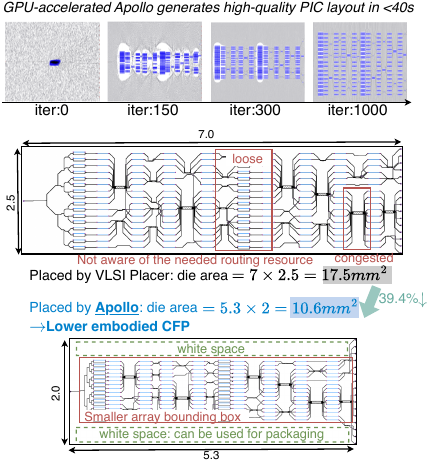}
    \caption{
    Compared to VLSI placer DREAMPlace~\cite{PLACE_TCAD2020_Lin}, \texttt{Apollo}~\cite{PLACE_ICCAD2025_Zhou} places the same photonic circuits in a $\sim$40\% smaller die with even larger whitespace left for packaging usage, further improving the PIC sustainability.}
    \label{fig:epda_layout}
    \vspace{-10pt}
\end{figure}

An automated, PIC layout automation engine would not only streamline the design process but also establish the foundation for \textbf{sustainable and scalable EPIC AI system design}. 
Here, we use PIC placer \texttt{Apollo}~\cite{PLACE_ICCAD2025_Zhou} and PIC router \texttt{LiDAR}~\cite{LiDAR_ISPD_Zhou, LiDAR2_ARXIV_Zhou} as examples to show how routing-informed placement and curvy-aware waveguide routing can deliver routable, compact, and robust layouts within minutes.
These tools show that automated optimization can achieve what manual design cannot: \textit{scalable exploration of the carbon–performance design space, leading to both faster development cycles and significantly more sustainable photonic hardware}.

\nameplacer is a GPU-accelerated, waveguide-routing-informed analytical placer specifically tailored for PICs. 
It generates highly routable photonic placement solutions by considering photonic-specific constraints during placement optimization, while \namerouter is a curvy-aware waveguide detailed router based on A$^\ast$ search. It generates waveguide routes that respect the minimum bending radius constraints and support dynamic waveguide crossing insertion. 
A \emph{central difficulty} in PIC layout is that \textbf{naïve area minimization is counterproductive}. 
Minimizing wirelength alone may yield compact placements, but often results in severe routing congestion. 
In photonics, bends and crossings consume significant area, and fixed port orientations introduce access challenges.
As a result, the most compact layout is rarely routable or robust due to severe crosstalk among nearby photonic devices.

To overcome this while maintaining chip compactness, \nameplacer incorporates waveguide-routing awareness directly into placement.
\ding{202}~Its bending-aware cost function (\textit{cosWA}) penalizes placements that would otherwise require excessive bends due to port misalignment, thereby improving routability and avoiding wasted area.
\ding{203}~Second, \nameplacer incorporates a net-spacing model to estimate the waveguide connections required between ports. By predicting potential crossings and reserving the space needed for their insertion, this model improves the routability of subsequent routing while avoiding excessive overestimation that would lead to area inefficiency.  
\ding{204}~Finally, \nameplacer enforces physical constraints, e.g., alignment and crosstalk-aware spacing, which are critical to circuit performance and robustness. 
After the placement stage, \texttt{LiDAR} performs automated curvy-aware detailed waveguide routing, generating feasible waveguide paths that honor minimum bending radius and dynamically inserting crossings when needed.

\begin{table*}[]
\centering
\caption{Device parameters used for CFP calculation of conventional and advanced EICs and SoTA EPICs}
\vspace{-5pt}
\resizebox{2\columnwidth}{!}{%
\begin{tabular}{|c|cc|cc|cc|c|c|}
\hline
EICs                   & \multicolumn{2}{c|}{Die Area ($mm^2$)}               & \multicolumn{2}{c|}{Power (W)} & \multicolumn{2}{c|}{Tech Node (nm)}     & TOPS     & Lifetime (Years) \\ \hline
Eyeriss v1~\cite{Chen2017EyerissJSSC}           & \multicolumn{2}{c|}{12.25}                                          & \multicolumn{2}{c|}{0.28}     & \multicolumn{2}{c|}{65}                 & 0.084    & 5                \\
Eyeriss v2~\cite{Chen2019EyerissV2}             & \multicolumn{2}{c|}{14.25}                                          & \multicolumn{2}{c|}{0.42}     & \multicolumn{2}{c|}{65}                 & 0.095    & 5                \\
PhyCU~\cite{Ju2024PhyCU}                  & \multicolumn{2}{c|}{3.80}                                            & \multicolumn{2}{c|}{0.15}     & \multicolumn{2}{c|}{28}                 & 0.336    & 5                \\
TranCIM~\cite{Tu2023TranCIM}                & \multicolumn{2}{c|}{6.82}                                           & \multicolumn{2}{c|}{0.09}     & \multicolumn{2}{c|}{28}                 & 1.48     & 7                \\
TPU v4~\cite{TPUV4}                 & \multicolumn{2}{c|}{370}                                            & \multicolumn{2}{c|}{200}       & \multicolumn{2}{c|}{7}                  & 275      & 7                \\
H100 SXM~\cite{NVIDIA2022H100Whitepaper}~\cite{NVIDIA2022H100PCIeBrief}               & \multicolumn{2}{c|}{814}                                            & \multicolumn{2}{c|}{700}       & \multicolumn{2}{c|}{5}                  & 3958     & 7                \\ \hline
EPICs                  & EIC Area ($mm^2$) & PIC Area ($mm^2$) & EIC Power (W)  & PIC Power (W) & EIC Tech Node (nm) & PIC Tech Node (nm) & TOPS     & Lifetime (Years) \\ \hline
Lightening Transformer~\cite{NP_HPCA2024_Zhu} & 17.60                           & 28.01                         & 9.79          & 7.30         & 28                 & 193                & 138.24   & 7                \\
TeMPO~\cite{NP_JAP2024_Zhang}                  & 23.91                           & 258.48                          & 9.79          & 7.30         & 45                 & 193                & 368.64   & 7                \\
CHORD w/o Reconfig.~\cite{NP_DAC25_Yin}    & 14.56                          & 0.04                       & 4.730          & 0             & 45                 & 193                & 20971.52 & 0.2              \\
CHORD w/ Opt.~\cite{NP_DAC25_Yin}          & 18.26                      & 3.74                       & 4.7978         & 0.05       & 45                 & 193                & 20976.64 & 5                \\
SCATTER w/o Opt.~\cite{NP_ICCAD2024_Yin}       & 3.56                            & 19.14                           & 13.18        & 10.13        & 45                 & 193                & 40.96    & 5                \\
SCATTER w/ Opt.~\cite{NP_ICCAD2024_Yin}         & 1.59                            & 10.81                           & 1.22          & 1.96         & 45                 & 193                & 40.96    & 5                \\ \hline
\end{tabular}
}
\label{tab:CFPparameters}
\vspace{-10pt}
\end{table*}

\noindent\textit{\underline{Insight}}.~
\nameplacer and \namerouter form a complete PIC layout synthesis flow that can generate compact yet routable layouts, reducing design time by orders of magnitude from days to minutes compared to manual flows, as illustrated in Fig.~\ref{fig:epda_layout}.
More importantly, the automated layout optimization directly reduces chip area and routing layer usage, lowering PIC manufacturing cost and the embodied CFP.
Beyond this, this fast, high-quality PIC layout toolkit further enables rapid PIC physical implementation with accurate estimation of key factors affecting the chip CFP, such as chip area, routing resource demand, and analog circuit robustness.
This capability enables designers to focus on end-to-end physical-aware system optimization with fast CFP-centric design space exploration.

\vspace{-3pt}
\section{Carbon-Aware Performance Analysis}
\vspace{-5pt}

Conventional CFP does not inform hardware selection: two platforms with similar carbon totals can differ by orders of magnitude in usable performance. Guided by ECO-CHIP~\cite{sudarshan2023ecochip}, we therefore separate total CFP into embodied and operational contributions and normalize by delivered performance, producing carbon-aware efficiency measures (e.g., CFP per task and performance per $kgCO_2e$) that are directly comparable across devices and workloads.

\vspace{-5pt}
\subsection{Carbon Footprint Modeling}
We adopt the standard total-CFP split:
\vspace{-5pt}
\begin{equation}
    \label{eq:ctot}
    C_{\mathrm{tot}} \;=\; C_{\mathrm{emb}} \;+\; \text{lifetime}\times C_{\mathrm{op}}.
    \vspace{-5pt}
\end{equation}

\paragraph{Embodied carbon ($C_{\mathrm{emb}}$)}
We express embodied CFP as the sum of manufacturing ($C_{\mathrm{mfg}}$), design ($C_{\mathrm{des}}$), and (if applicable) advanced packaging/heterogeneous-integration ($C_{\mathrm{HI}}$):
\vspace{-5pt}
\begin{equation}
    \label{eq:cemb}
    C_{\mathrm{emb}} \;=\; C_{\mathrm{mfg}} \;+\; C_{\mathrm{des}} \;+\; C_{\mathrm{HI}}.
    \vspace{-5pt}
\end{equation}
Design CFP is amortized over volume, and chiplet/package reuse:
\vspace{-8pt}
\begin{equation}
    \label{eq:cdes}
    C_{\mathrm{des}} \;=\; \sum_{i}C_{\mathrm{des},i}/N_{M,i} \;+\; C_{\mathrm{des,\,comm}}/N_{S},
    \vspace{-5pt}
\end{equation}
where $N_{M,i}$ is the number of manufactured units of chiplet $i$ and $N_S$ is the number of systems that share the common design/communication IP.

\paragraph{Yield- and node-aware manufacturing}
We follow ECO-CHIP’s node- and yield-aware treatment of manufacturing CFP. 
In short, die area scales with node- and block-specific density $D_T(d,p)$, and die yield $Y(d,p)$ is modeled by a negative-binomial function of defect density $D_0(p)$ and clustering $\alpha$. 
The manufacturing carbon per-unit area (CFPA) is divided by $Y(d,p)$, so larger dies or immature nodes (higher $D_0$) incur higher embodied CFP. 
We also account for the carbon impact of the unusable silicon at the wafer's circular edge via dies-per-wafer (DPW) and amortizing its cost. %

\paragraph{Operational carbon ($C_{\mathrm{op}}$)}
Operational CFP multiplies energy-at-use ($E_{\mathrm{use}}$) by the carbon intensity ($CI_{\mathrm{use}}$) of the electricity mix: $C_{\mathrm{op}} \;=\; \mathrm{CI}_{\mathrm{use}} \times E_{\mathrm{use}}$.
At the device level, we model the average energy during active operation over its duty cycle:
\vspace{-5pt}
\begin{equation}
    \small
    \label{eq:euse}
    E_{\mathrm{use}} \;=\; P_{\mathrm{active}}\times t_{\mathrm{on}} \;\;\;\; \text{with} \;\;\; P_{\mathrm{active}}=\frac{\mathrm{workload\ ops}}{\mathrm{TOPS/W}}\; .
    \vspace{-5pt}
\end{equation}
(For CMOS devices, $P_{\mathrm{active}}$ can be broken down into dynamic and leakage components; for EPICs, non-passive overheads such as DAC/ADC, laser/driver, and thermal control should be included explicitly.)

\subsection{Why CFP Alone Is Not Sufficient}
The same embodied or operational CFP can correspond to radically different delivered performance and service levels. To compare systems fairly, carbon should be normalized by \emph{what the device delivers}: throughput, latency, and utilization. Inspired by prior work~\cite{NP_IEEE_Sudarshan2024}, we introduce two carbon-aware performance metrics.

\vspace{-5pt}
\subsection{Carbon-Normalized Metrics}
\paragraph{Carbon per task ($CFP_{task}$).}
For real-time inference or simulation, a direct measure is \emph{carbon per completed task}:
\begin{equation}
\vspace{-5pt}
    \small
    \label{eq:cfp_per_task}
    \mathrm{CFP_{task}} \;=\; C_{\mathrm{emb}}/N_{\mathrm{life}} \;+\; \mathrm{CI}_{\mathrm{use}}\times P_{\mathrm{active}}/\mathrm{\Theta},
\end{equation}
where $\Theta$ is the throughput (tasks/s) at a service-level agreement (SLA) operating point, $N_{\mathrm{life}}$ is the lifetime number of tasks (e.g., frames, inferences) delivered by the device. 
In practice, we report \emph{kgCO$_2$e per inference} (or per frame) at a specified SLA (batch size and sequence length).

\paragraph{Performance per carbon footprint (Perf/CFP)}
To compare devices across broader operating points, we also report \emph{performance per kgCO$_2$e}:
\begin{equation}
\small
    \label{eq:perf_per_cfp}
    \mathrm{Perf/CFP} \;=\; \mathrm{Perf}/(C_{\mathrm{emb}} \;+\; \mathrm{CI}_{\mathrm{use}}\times E_{\mathrm{life}}).
\end{equation}
$\mathrm{Perf}$ can be TOPS, tasks/s (e.g., FPS, inferences/s), or problem-specific throughput. 

\begin{table*}[t]
\centering
\caption{Comparison of the chip area ($mm^2$), routability, runtime, and routed wirelength of the max insertion loss path.}
\vspace{-5pt}
\renewcommand{\arraystretch}{1.2}
\resizebox{\textwidth}{!}{
\begin{tabular}{|c|cccc|cccc|cccc|}
\hline
\multirow{2}{*}{Benchmark} & \multicolumn{4}{c|}{Semi-automation} & \multicolumn{4}{c|}{DREAMPlace$^\ast$~\cite{PLACE_TCAD2020_Lin}} & \multicolumn{4}{c|}{\nameplacer} \\ \cline{2-13} 
& area ($mm^2$) & Routability & Runtime & WL ($mm$) & area ($mm^2$) & Routability & Runtime (s) & WL ($mm$) & area ($mm^2$) & Routability & Runtime (s) & WL ($mm$) \\ \hline
Clements\_8$\times$8 & 3.5 & 100\% & ($\sim$12h)+10s & 1.91 & 7.7 & 100\% & 41 & 3.46 & 3.3 & 100\% & 31 & 1.83 \\
Clements\_16$\times$16 & 13.6 & 100\% & ($\sim$12h)+92s & 3.56 & 22.6 & 100\% & 135 & 5.14 & 12.7 & 100\% & 117 & 3.42 \\
ADEPT\_8$\times$8 & 4.4 & 100\% & ($\sim$1day)+49s & 3.73 & 5.4 & 100\% & 61 & 3.88 & 3.5 & 100\% & 31 & 2.46 \\
ADEPT\_16$\times$16 & 12.2 & 100\% & ($\sim$1day)+168s & 8.30 & 17.5 & 98.24\% & 223 & 7.39 & 10.6 & 99.68\% & 95 & 5.55 \\
ADEPT\_32$\times$32 & 31.2 & 99.60\% & ($\sim$1day)+451s & 9.59 & 34.0 & 97.95\% & 396 & 9.57 & 27.2 & 99.75\% & 306 & 7.86 \\
ADEPT\_64$\times$64 & 90.0 & 99.02\% & ($\sim$1day)+1451s & 16.18 & 77.2 & 95.74\% & 1557 & 13.26 & 69.7 & 98.66\% & 822 & 11.49 \\ \hline
Geo-mean & 25.8 & 99.77\% & $\sim$16h & 7.21 & 27.4 & 98.65\% & 402 & 7.12 & 21.2 & 99.68\% & 234 & 5.44 \\
Ratio & 1.00 & 1.00 & 1.00 & 1.00 & 1.06 & 0.98 & 0.006 & 0.98 & 0.82 & 0.99 & 0.004 & 0.75 \\ \hline
\end{tabular}
}
\label{tab:pr_result}
\vspace{-13pt}
\end{table*}

Table \ref{tab:CFPparameters} summarizes the parameters we use in the carbon model: die area, active power at the SLA operating point, technology node(s), sustained performance (TOPS), and the service lifetime used to amortize embodied carbon. 
For EPICs we report the EIC and PIC footprints and powers separately (EIC area/power and node for drivers, DAC/ADC, digital control; PIC area/power and node for the integrated photonics), with the PIC node set to AIM Photonics’ 193 nm process for all integrated photonic designs in this study; electronic nodes are taken from the respective papers. 
Lifetimes are chosen to reflect typical deployment horizons used in prior carbon-aware hardware analyses and industry practice: we assign 5 years to convolution-only ASICs (research/edge designs with shorter refresh cycles and more rapid obsolescence), and 7 years to Transformer-capable accelerators (advanced GPU/TPU and EPIC systems) to reflect longer datacenter depreciation, multi-generation model reuse, and higher reuse across workloads. 
These lifetimes are only used to amortize embodied CFP; the operational CFP is computed from measured active power and sustained throughput at the stated operating point.

\begin{figure}
    \centering
    \vspace{-5pt}
    \includegraphics[width=0.85\columnwidth]{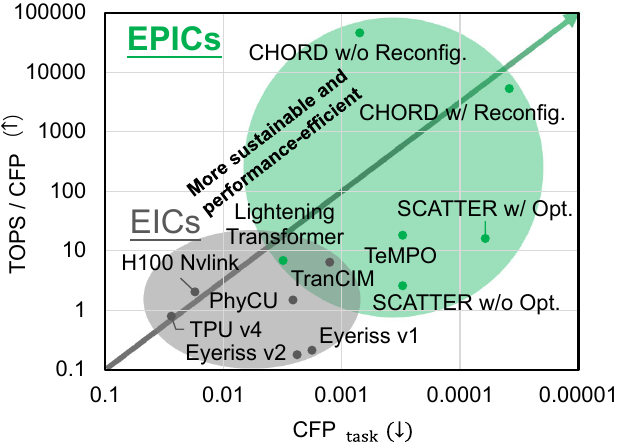}
    \vspace{-4pt}
    \caption{Carbon-performance tradeoff across devices.
    The y-axis measures performance per carbon footprint, where we take TOPS as the performance metric (TOPS/CFP). Devices toward the upper-right corner represent the most sustainable and performance-efficient designs.}
    \label{fig:cfpcomp}
    \vspace{-18pt}
\end{figure}

Figure~\ref{fig:cfpcomp} shows that EPIC designs consistently occupy the upper-right region of the tradeoff curve, demonstrating both reduced carbon footprint per task and higher performance per unit carbon. This emphasizes the natural sustainability advantage of combining photonics with electronics.

As stated in Sec.~\ref{subsec:tsevvar} and~\ref{subsec:tsevro}, EPIC designs benefit strongly from cross-layer optimization. 
The co-design of algorithms, circuits, and architecture allows the die size to be reduced while minimizing electrical–optical conversion overhead.
With in-situ light redistribution and input gating, the optimized SCATTER system reuses optical power more effectively, achieves higher optical SNR, and lowers active power consumption. 
These improvements shift the optimized SCATTER design markedly upward and to the right compared to its unoptimized baseline, underscoring how efficiency-driven design directly translates into better carbon amortization.

Likewise, CHORD illustrates the sustainability benefits of reconfigurability. 
By enabling multiple degrees of freedom through system-level reconfigurability, CHORD avoids the obsolescence of fixed diffractive systems. Its longer functional lifetime can amortize embodied carbon across many more tasks and application generations, driving its $CFP_{task}$ lower while maintaining high performance per carbon.

\vspace{-3pt}
\section{Evaluation of Automated PIC Layout Engine}
\vspace{-3pt}
We evaluate the capability of \nameplacer and \namerouter to generate highly compact layouts that minimize chip area while ensuring that more than 98\% of nets can be successfully routed.

\noindent\textbf{Benchmarks}. We adopt representative benchmarks from prior work~\cite{PLACE_ICCAD2025_Zhou}, including Clements-style MZI arrays~\cite{NP_NATURE2017_Shen} and an auto-searched photonic tensor core design (ADEPT)~\cite{NP_DAC2022_Gu}. 
These benchmarks cover circuits of varying scales. 

\noindent\textbf{Baselines}.~We compare \nameplacer against two baselines. 
The first is DREAMPlace$^\ast$, a variant of DREAMPlace~\cite{PLACE_TCAD2020_Lin} in which the original linear wirelength model often produces highly unroutable solutions. 
To mitigate this, a quadratic wirelength is used, and the chip area and target density are adjusted to ensure sufficient spacing between components for routing. 
The second is a semi-automation method using manual placement scripts. 
The spacing between components is estimated manually, and \namerouter is employed for routing.

Table~\ref{tab:pr_result} summarizes the comparison between \nameplacer and the baselines in terms of area, routability, runtime, and critical-path wirelength (WL). 
Compared with the semi-automation approach, \nameplacer achieves on average an 18\% reduction in chip area while maintaining comparable routability, and also reduces WL by 25\%. 
These results demonstrate the strong capability of \nameplacer to generate compact placement solutions. 
It is worth noting that although the semi-automation method allows script reuse across circuits of the same type, developing such scripts is highly time-consuming: approximately one day was required for \texttt{ADEPT} and about 12 hours for \texttt{Clements}. 
In contrast, \nameplacer only requires an initial random placement of components near the layout center; during the subsequent optimization process, the components are automatically spread out while reserving sufficient routing resources in congested regions.
Moreover, the semi-automation method also cannot guarantee routability for all nets, largely due to the extensive crossings in \texttt{ADEPT}, which introduce significant routing challenges. 
Nevertheless, in such cases, local manual adjustments to the failed nets can still be applied to save considerable effort. 
For the \texttt{Clements} benchmarks, however, all methods were able to achieve full routability. Regarding the DREAMPlace$^\ast$, which lacks a model for waveguide routing resources, exhibits lower routability on \texttt{ADEPT} even when its die size is increased to about 1.3$\times$ that of \nameplacer.

Overall, these results highlight the effectiveness and practicality of \nameplacer in producing compact, routable, and efficient placement solutions for large-scale PIC layouts.

\vspace{-5pt}
\section{Conclusion}
The rise of AI has forced computing to confront its carbon cost head-on. 
Electronic–photonic integrated circuits provide an opportunity to rethink this trajectory: light delivers speed and efficiency that electronics alone cannot, while careful co-design and automation turn those raw advantages into lasting sustainability.
What emerges is a simple but powerful principle: compact, reconfigurable, and enduring hardware directly translates into lower embodied and operational carbon.
The path forward is not incremental scaling, but a fundamental new design mindset with carbon awareness at every layer: devices, circuits, architectures, and design automation working in concert.
If pursued deliberately, EPICs will not just accelerate tomorrow’s AI, but also set a precedent for how to build lifelong, carbon-conscious computing systems.


% \bibliographystyle{IEEEtran} %
% \bibliography{./ref/Top_sim,./ref/NN,./ref/NP,./ref/ALG, ./ref/DFM, ./ref/PD, ./ref/vidya, ./ref/IEEESettings}

\end{document}